\documentclass[11pt,a4paper]{article}

\usepackage{fullpage}
\usepackage{amsmath}
\usepackage{amssymb}
\usepackage{amsthm}
\usepackage{mathrsfs}
\usepackage{graphicx}
\usepackage{psfrag}
\usepackage{citesort}

\newcommand {\be} {\begin{eqnarray*}}
\newcommand {\ee} {\end{eqnarray*}}
\newcommand {\bea} {\begin{eqnarray}}
\newcommand {\eea} {\end{eqnarray}}

\newcommand{\qedd}{\hfill $\blacksquare$}

\title{Manifold invariants affect dynamics in ADS gravity}
\author{\textbf{Tom$\acute{\mbox{a}}\check{\mbox{s}}$ Liko}\footnote{Electronic mail: tliko@math.ualberta.ca}\\
\\{\small \it Department of Mathematical and Statistical Sciences}\\
{\small \it University of Alberta}\\
{\small \it Edmonton, AB T6G 2G1, Canada}}

\begin{document}

\maketitle

\begin{abstract}

The first-order Holst action with negative cosmological constant is rendered finite by requiring functional
differentiability on the configuration space of tetrads and connections.  The surface terms that arise in
the action for ADS gravity are equivalent to the Euler and Pontryagin densities with fixed weight factors;
these terms modify the Noether charges that arise from diffeomorphism invariance of the action.

\end{abstract}

\hspace{0.35cm}{\small \textbf{PACS}: 04.20.Cv; 04.20.Fy; 04.70.Dy}

Let $\mathcal{M}$ be a four-dimensional manifold with boundary $\partial\mathcal{M}$.  The configuration
space $\mathscr{C}$ of the gravitational field is the pair $\{e,A\}$, consisting of the coframe
$e^{I}:=e_{a}^{\phantom{a}I}dx^{a}$ and spin connection $A^{IJ}:=A_{a}^{\phantom{a}IJ}dx^{a}$ (with spacetime indices
$a,b,\ldots\in\{0,\ldots,3\}$ and internal indices $I,J,\ldots\in\{0,\ldots,3\}$).  In terms of $e$ and $A$,
the spacetime metric and curvature $2$-form are (resp.) $g_{ab}=\eta_{IJ}e_{a}^{\phantom{a}I} \otimes e_{b}^{\phantom{a}J}$
and $\Omega_{\phantom{a}J}^{I}=dA_{\phantom{a}J}^{I}+A_{\phantom{a}K}^{I} \wedge A_{\phantom{a}J}^{K}=(1/2)R_{\phantom{a}JKL}^{I}e^{K}
\wedge e^{L}$, with $R_{\phantom{a}JKL}^{I}$ being the Riemann tensor.

The first-order Holst action with negative cosmological constant $\Lambda:=-3\ell^{-2}\in\mathbb{R}_{-}$
and Barbero-Immirzi parameter $\gamma\in\mathbb{R}\setminus\{0\}$ on a four-dimensional manifold
$(\mathcal{M},e,A)$ with generic boundary $\partial\mathcal{M}$ is \cite{acotz,durkow,durka,liko1}:
\bea
I &=& \frac{1}{32\pi}\int_{\mathcal{M}}\epsilon_{IJKL}\left(e^{I} \wedge e^{J} \wedge \Omega^{KL}
      + \frac{1}{2\ell^{2}}e^{I} \wedge e^{J} \wedge e^{K} \wedge e^{L} + \frac{\ell^{2}}{2}\Omega^{IJ} \wedge \Omega^{KL}\right)\nonumber\\
  & & - \frac{1}{\gamma}\left(e_{I} \wedge e_{J} \wedge \Omega^{IJ} + \ell^{2}\Omega_{IJ} \wedge \Omega^{IJ}\right) \; .
\label{foaction}
\eea
This form of the action can be used to derive the Noether charges associated with diffeomorphisms.  Specifically,
for a smooth vector field $\xi$ generating a spacetime diffeomorphism, the symplectic current 3-form is
$\star \mathcal{J}=\Theta(e,A,\delta A)-I_{\xi}\mathcal{L}$, with $\Theta$ being the (first) variation of the surface
terms in (\ref{foaction}), $\mathcal{L}$ the Lagrangian density in (\ref{foaction}), $\star$ denoting Hodge duality
and $I_{\xi}$ denoting the contraction operator acting on $p$-forms.  For the Holst-ADS action, the symplectic current
is a total derivative, and can be integrated over a two-dimensional spatial cross section $\Sigma$ of $\partial\mathcal{M}$.
The resulting conserved charge that arises from diffeomorphism invariance of the first-order Holst-ADS action is:
\bea
\mathcal{Q}[\xi] = \frac{\ell^{2}}{32\pi}\oint_{\Sigma}\epsilon_{IJKL}\bar{\Omega}^{IJ} \wedge I_{\xi}A^{KL}
         - \frac{1}{\gamma}\bar{\Omega}_{IJ} \wedge I_{\xi}A^{IJ} \, ,
\label{adscharge}
\eea
with $\bar{\Omega}:=\Omega+(1/\ell^{2})e\wedge e$.  This charge is an extension of the Noether charge found by Aros
\emph{et al} \cite{acotz} to include a real and non-zero Barbero-Immirzi parameter.  The presence of $\gamma$ in the
action therefore changes the Noether charge for ADS spacetimes.  This contribution to the charge vanishes for spacetimes
that have diagonalizable metrics or zero Pontryagin number \cite{durkow,durka,liko1}.

\emph{\bf Example 1: Euler density and black-hole entropy.}
In this example, we consider spacetimes containing a (local) Killing horizon $\Delta$ as inner boundary and an
asymptotic region $\mathscr{I}$ as outer boundary.  The topology of $\Delta$ is $\mathbb{R}\times\Sigma_{g}$,
with $\Sigma_{g}$ being a compact 2-manifold with genus $g\in\mathbb{Z}_{+}\cup\{0\}$.  The class of spacetimes
thus include the Schwarzschild-ADS and Kerr-ADS families of solutions describing black holes with `topological'
horizons (see, e.g. Galloway \emph{et al} \cite{gsww} and references therein).

For the class of spacetimes under consideration, the Euler term is non-zero and the Pontryagin term is zero
in the \emph{on-shell} first-order action.  A non-zero Euler term in the action modifies the entropy of the
corresponding black holes.  The entropy for these spacetimes that satisfies the first law of black-hole
mechanics is $\mathscr{S}:=\mathscr{A}/4+\pi\ell^{2}(1-g)$, with $\mathscr{A}$ being the surface area of the
two-dimensional cross section $\Sigma_{g}$ of the horizon \cite{jacmye,liko2}.  The toroidal horizon is a
special case, for which the entropy is $\mathscr{S}=\mathscr{A}/4$.

Let us consider the merging of two black holes.  The entropies of these black holes are $\mathscr{S}_{1}=(1/4)
[\mathscr{A}_{1}+4\pi\ell^{2}(1-g_{1})]$ and $\mathscr{S}_{2}=(1/4)[\mathscr{A}_{2}+4\pi\ell^{2}(1-g_{2})]$.
If the total entropy before the black holes merge is
$\mathscr{S}=\mathscr{S}_{1}+\mathscr{S}_{2}=(1/4)[\mathscr{A}_{1}+\mathscr{A}_{2}+4\pi\ell^{2}(2-g_{1}-g_{2})]$,
and the total entropy of the final black hole configuration is
$\mathscr{S}^{\prime}=(1/4)[\mathscr{A}^{\prime}+4\pi\ell^{2}(1-g^{\prime})]$, the second law of thermodynamics will
hold if and only if $\mathscr{S}^{\prime}>\mathscr{S}_{1}+\mathscr{S}_{2}$.  Thus we have the following upper
bound on the ADS radius $\ell$:
\bea
\ell < \left[\left(\frac{1}{4\pi}\right)
       \left(\frac{\mathscr{A}^{\prime}-\mathscr{A}_{1}-\mathscr{A}_{2}}{1+g^{\prime}-g_{1}-g_{2}}\right)\right]^{1/2} \; .
\label{bound}
\eea
The reality of $\ell$ implies that $\mathscr{A}^{\prime}>\mathscr{A}_{1}+\mathscr{A}_{2}$ \emph{if and only if}
$g^{\prime}>g_{1}+g_{2}-1$.  In other words, the number of genera cannot decrease in a physical process for which
the second law of thermodynamics (in the form $\mathscr{S}^{\prime}>\mathscr{S}_{1}+\mathscr{S}_{2}$) \emph{and}
the second law of black-hole mechanics (in the form $\mathscr{A}^{\prime}>\mathscr{A}_{1}+\mathscr{A}_{2}$) hold
together.  Thus, for example, two toroidal black holes cannot merge into a final toroidal configuration because
the number of genera of the final state has to be at least $2$.
\qedd

\emph{\bf Remarks.}
({\bf i}) The bound (\ref{bound}) is consistent with area bounds that were found by Gibbons \cite{gibbons} and
by Woolgar \cite{woolgar}.  For black holes in ADS spacetime, the surface area of the black hole is constrained
by the bound: $\mathscr{A}\geq4\pi(g-1)/|\Lambda|$.  This bound says that higher-genus horizons necessarily have
greater areas.  ({\bf ii}) Similar area bounds can be obtained for black holes in spacetimes with \emph{positive}
cosmological constant.  For black holes merging in de Sitter spacetimes, it turns out that the second law of
thermodynamics holds while the second law of black-hole mechanics is \emph{violated}.

\emph{\bf Example 2: Pontryagin density and gravitational entropy.}
In this example, we consider a more general class of spacetimes than the ones considered in Example 1.  For these
spacetimes, the Pontryagin number is taken to be non-zero.  Spacetimes with non-zero Pontryagin number include the
Taub-NUT solution \cite{hawking}.  The entropy of the Taub-NUT-ADS spacetime is found to be \cite{durka,liko1}
\bea
\mathscr{S} = \mathscr{S}_{0} + \frac{4\pi N^{2}}{\gamma}\left(1 - \frac{N^{2}}{\ell^{2}}\right) \; .
\eea
Here, $\mathscr{S}_{0}$ is the entropy of the Taub-NUT-ADS solution without the Holst term \cite{mann,cejm,acotz}.
\qedd

\emph{Summary.}
Topological invariants of a manifold with boundary, although total derivatives in the action, affect the dynamics
of black objects in four dimensions.

\emph{Acknowledgements.}
The author thanks E. Woolgar for discussions and R. Durka for checking an
earlier draft of this proceeding.  The work presented here was supported by NSERC.

\end{document}